\documentclass{emulateapj}

\newcommand{\msun}{M$_{\sun}$}
\newcommand{\kms}{km s$^{-1}$}
\newcommand{\msuns}{M$_{\sun}~$}
\newcommand{\kmss}{km s$^{-1}~$}

\shorttitle{Encounters with Massive Star-Disk Systems}

\begin{document}

\title{Stellar Encounters with Massive Star-Disk Systems}

\author{Nickolas Moeckel and John Bally} 
\affil{ Center for Astrophysics and Space Astronomy, \\
        Department of Astrophysical and Planetary Sciences \\
	University of Colorado, 389 UCB, Boulder, CO 80309-0389}
\email{moeckel@colorado.edu}

\begin{abstract}
The dense, clustered environment in which massive stars form can lead to interactions with neighboring stars.  It has been hypothesized that collisions and mergers may contribute to the growth of the most massive stars.  In this paper we extend the study of star-disk interactions to explore encounters between a massive protostar and a less massive cluster sibling using the publicly available SPH code GADGET-2.
Collisions do not occur in the parameter space studied, but the end state of many encounters is an eccentric binary with a semi-major axis $\sim 100$ AU.  Disk material is sometimes captured by the impactor.  Most encounters result in disruption and destruction of the initial disk, and periodic torquing of the remnant disk.  We consider the effect of the changing orientation of the disk on an accretion driven jet, and the evolution of the systems in the presence of on-going accretion from the parent core.  
\end{abstract}


\section{INTRODUCTION}
In the standard picture of star formation, stars form from the collapse and fragmentation of turbulent molecular cloud cores on a timescale of $\sim 10^{5}$ yr for individual stars to $\sim 10^{7}$ yr for large OB associations.  Accretion onto the prestellar core from the surrounding envelope can occur spherically, or if the material has some angular momentum, from the circumstellar disk.  While the details of these processes are becoming understood for isolated low mass stars ($M_{\star}\lesssim10$ \msun), the formation of clusters, multiple systems, and massive stars are less clear.  Simply extending the process for low mass stars to massive star formation presents some problems.  

The Jeans mass of cloud cores containing massive stars is $\sim0.3$ \msun, while observed stellar masses are $\sim50$ \msuns \citep{bon98a}.  \citet{wol87} argue that radiation pressure from the central star may halt or reverse spherical accretion for masses greater than about 20 \msun.  However, \citet{mck03} suggest that high external pressure can overcome this obstacle.  Calculations of accretion from the envelope onto a disk, and from the disk onto the star \citep{yor99,yor02} have resulted in protostars of $\sim30$\msun.  \citet{kru05d,kru05b} argue that anisotropies in the stellar radiation pressure and instabilities in the accretion flow can circumvent the radiation pressure problem.  However, these scenarios deal solely with the formation of a star in isolation.
  
Young massive stars are usually found in dense clusters \citep{cla00}, often near the central region \citep{hil98}, and have a high multiplicity fraction \citep{sta00, gar01}.  These factors suggest that interactions between neighboring stars are likely during cluster formation.  \citet{bon02} performed numerical simulations of accretion and merging in a cluster of 1000 identical mass stars.  They found that 19 mergers occurred, driven by the dynamics of small subgroups within the cluster.  Though the stellar densities they required ($\sim 10^8$ star pc$^{-3}$) are far in excess of those observed, it can be argued that such high densities exist briefly during the initial collapse of a high density molecular cloud core.  \citet{bon98a} and \citet{bon05} suggest a scenario wherein accretion onto a wide, low mass binary leads to a tight binary with a system mass of 30-50 \msun, which, with its larger cross section, could be driven to collision by an encounter with a third body.

If one considers a cluster with number density $10^6$ pc$^{-3}$, a velocity dispersion of 3 \kms, and populated by typical pre-main-sequence protostars with masses 0.3 \msuns and radii 0.05 AU, one can estimate the mean time between collisions \citep{bin87} as $t_{col} \sim4 \times 10^8$ yr.  This corresponds to a virialized cluster core with, e.g., 100 \msuns ($\sim$ 330 protostars) and radius $\sim$ 0.04 pc.  Even at such an extreme number density, this rate is not high enough for collisions to be a likely star formation mechanism.  \citet{bal05} suggest that circumstellar disks and envelopes could play a role beyond accretion by increasing the interaction radius, and therefore assisting stellar capture.  

The cross section of a star-disk system is much larger than the cross section of a single star, with disks commonly a few$\times10^2$ AU in radius.
Material ejected during an encounter carries away kinetic energy from the impactor, which expands the range of initial conditions that could lead to a merger.  Additionally, by sweeping through the disk the impactor is slowed both by accreting material, and by the gravitational effects of wakes and spiral arms induced in the disk.   
Can these effects increase the merger radius of the protostars to $\sim 100$ AU?  The increased gravitational focusing between a 20 \msuns star in a mass-segregated cluster with density $n_{\star} = 10^{5}$ pc$^{-3}$, similar to the density found in \citet{bon03}, can drive the timescale for encounters within 200 AU to less than $10^{5}$ yr; encounters with stars more massive than 1 \msuns take place on a timescale $\sim 10^{5}$ yr (Moeckel \& Bally 2006, in preparation).  Can the ensuing interactions lead to mergers, or to binaries which could evolve further through third body interactions or accretion onto the system?

Several authors have considered encounters between star-disk systems.  \citet{hel95}, \citet{lar97}, \citet{bof98}, and \citet{pfa05} used n-body and SPH codes, while \citet{hal96} and \citet{hal97} used a reduced-three body approach to study encounters ranging from penetrating to distant.  \citet{cla91} approached the question of binary formation due to star-disk encounters analytically, while \citet{ost94} analytically treated the case of more distant encounters.  \citet{wat98,wat98a} performed SPH simulations of encounters wherein both stars have circumstellar disks.  Of the numerical studies dealing with close encounters, none consider stellar mass ratios $m_{2}/m_{1} < 0.5$, and \citet{hal97} and \citet{pfa04} are the only studies that consider repeated encounters.
In this paper we present the results of numerical simulations concerning repeated, penetrating encounters between a massive protostar with a disk, and a less massive stellar impactor, to extend the parameter space considered in previous studies toward a regime suitable for the early life of a high mass star in a star forming cloud core.  The primary system is simulated in a late stage of formation, when the ratio of disk to stellar mass $M_d/M_{\star} = 0.1$, though we consider one case with an accreting envelope surrounding the system.

\section{SIMULATION METHOD}
The large dynamic and spatial ranges inherent to the problem, combined with the lack of a convenient fixed geometry, point naturally toward the use of a Lagrangian method. We use the publicly available smoothed particle hydrodynamics (SPH) code GADGET-2 \citep{spr05}.  The equations of motion integrated by GADGET-2 self-consistently include the so-called $\nabla h$ terms \citep{spr02,mon02}.  Rather than advancing the internal energy, an entropic function is integrated, from which the thermal energy can be evaluated.  Gravity is computed by means of a tree. 

GADGET-2 can accommodate a mix of collisional and collisionless particles.  We model the stars as point masses interacting only through gravity, while the disk particles are subject to both gravitational and gas dynamical forces.  We modified GADGET-2 so that stellar particles behave as sink particles \citep{bat95}, accreting gas that falls within a set accretion radius.  A particle is accreted only if it is energetically bound to the star, and if it has less angular momentum around the accretor than that of a Keplerian orbit at the accretion radius.  Following accretion, the gas particle is removed from the computation.  Its mass and linear momentum are transfered to the star, which is placed at the center of mass of the two particles.  

The accretion radius must satisfy two criteria.  First, it should be small compared to the gravitational radius, $r_{g} = \sqrt{2GM_{acc} / v^2}$, with $M_{acc}$ the mass of the accretor and $v$ its velocity through the gas.  \citet{ruf96} considered the effect of an accretor's physical size relative to its gravitational radius in Bondi-Hoyle accretion.  The physical radius of an accreting star is much less than it's gravitational radius; if $r_{acc} \gtrsim 0.2 r_{g}$, the linear momentum transfer from directly accreting upstream material (versus from a gravitationally focused accretion stream) changes sign.  The accreting spheres considered in \citet{ruf96} are totally absorbing; our energy and angular momentum tests help to some extent to decrease spurious accretion from upstream material.  Ideally, $r_{acc}$ could be set strictly according to the criterion $r_{acc}/r_{g} \le 0.2$.  However, a resolution limit also needs to be taken into account.  If the SPH smoothing lengths of the gas particles nearest the accretor are large enough that they are neighbors with particles on the other side of the star, then they can be artificially supported from accretion by unphysical gas pressure.  Likewise, if the gravitational smoothing length, which we set to be approximately equal to the average SPH smoothing length, is larger than $r_{acc}$ then accretion will be stunted.

  A 5 \msuns accreting object passing through a Keplerian disk around a 20 \msuns star with a closest approach $\sim50$ AU will have $r_{g} \gtrsim 25$ AU for a prograde encounter, and $r_{g} \gtrsim 5$ AU for a retrograde encounter.  In practice, for the prograde case and for inclinations such that the passage of the impactor through the disk is out of plane, $r_{acc} = 5$ AU is sufficiently small to avoid affecting the results.  In retrograde encounters, this is clearly too large as $r_{acc} \gtrsim r_{g}$.  An accretion radius less than 1 AU is suspect with the resolution we have.  Tests indicate that accretion radii in the range $0.75$ AU$< r_{acc} < 1.5$ AU produce comparable results; we choose $r_{acc} = 1.5$ AU for simulations with inclination angles less than 45 degrees from retrograde.

Since we are interested in encounters spanning multiple binary orbits, we require long term stability of the disks.  Because of the artificial viscosity employed in SPH, disks have a tendency to rapidly lose mass onto sink particles.  Following \citet{bat95}, we apply a corrective outward pressure force to particles close to the accretion radius to compensate for the lack of pressure support from particles interior to the disk inner edge.  This corrective force is only applied to particles in the disk around the primary star.
 \begin{figure*}
 \centering
  \plottwo{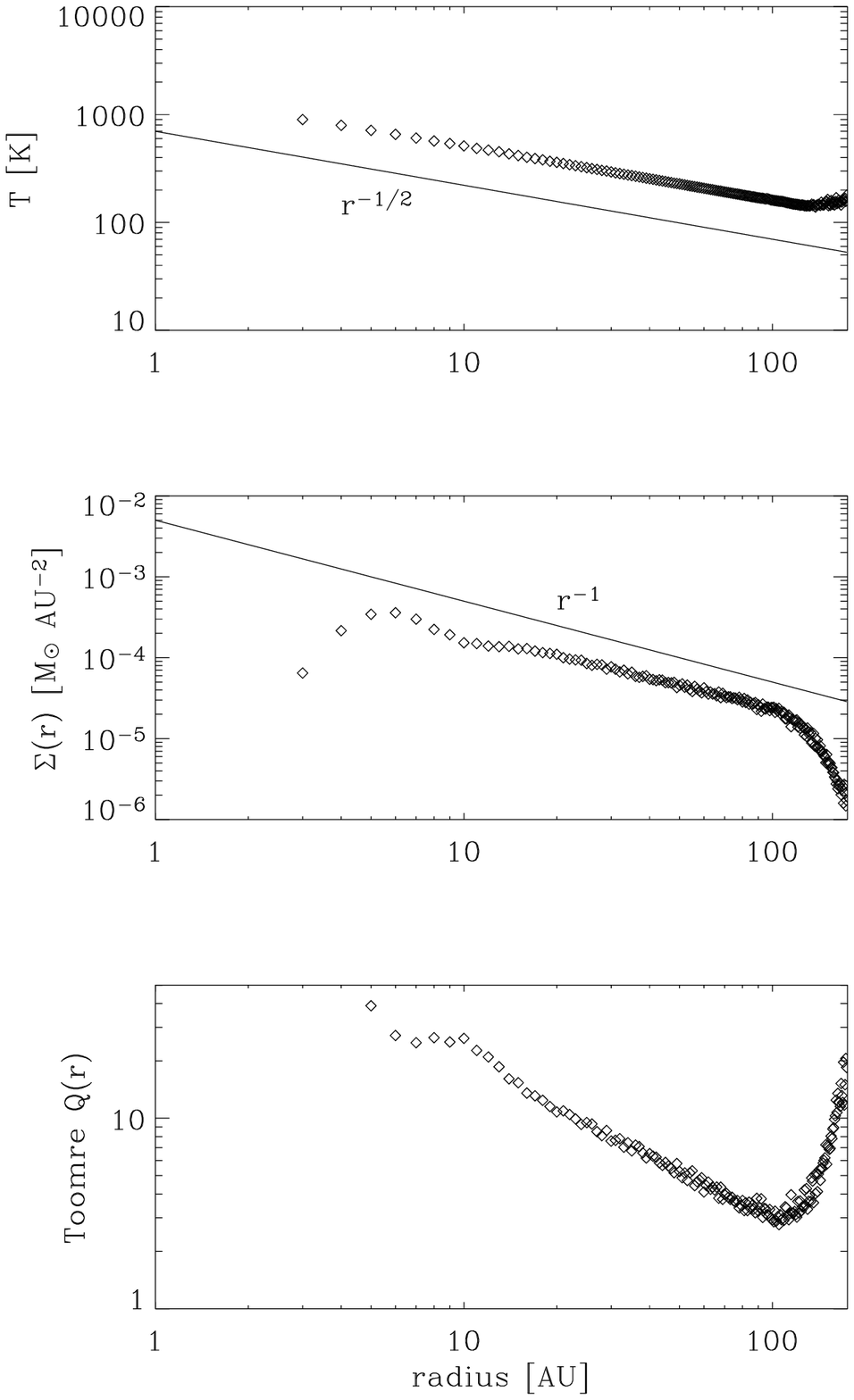}{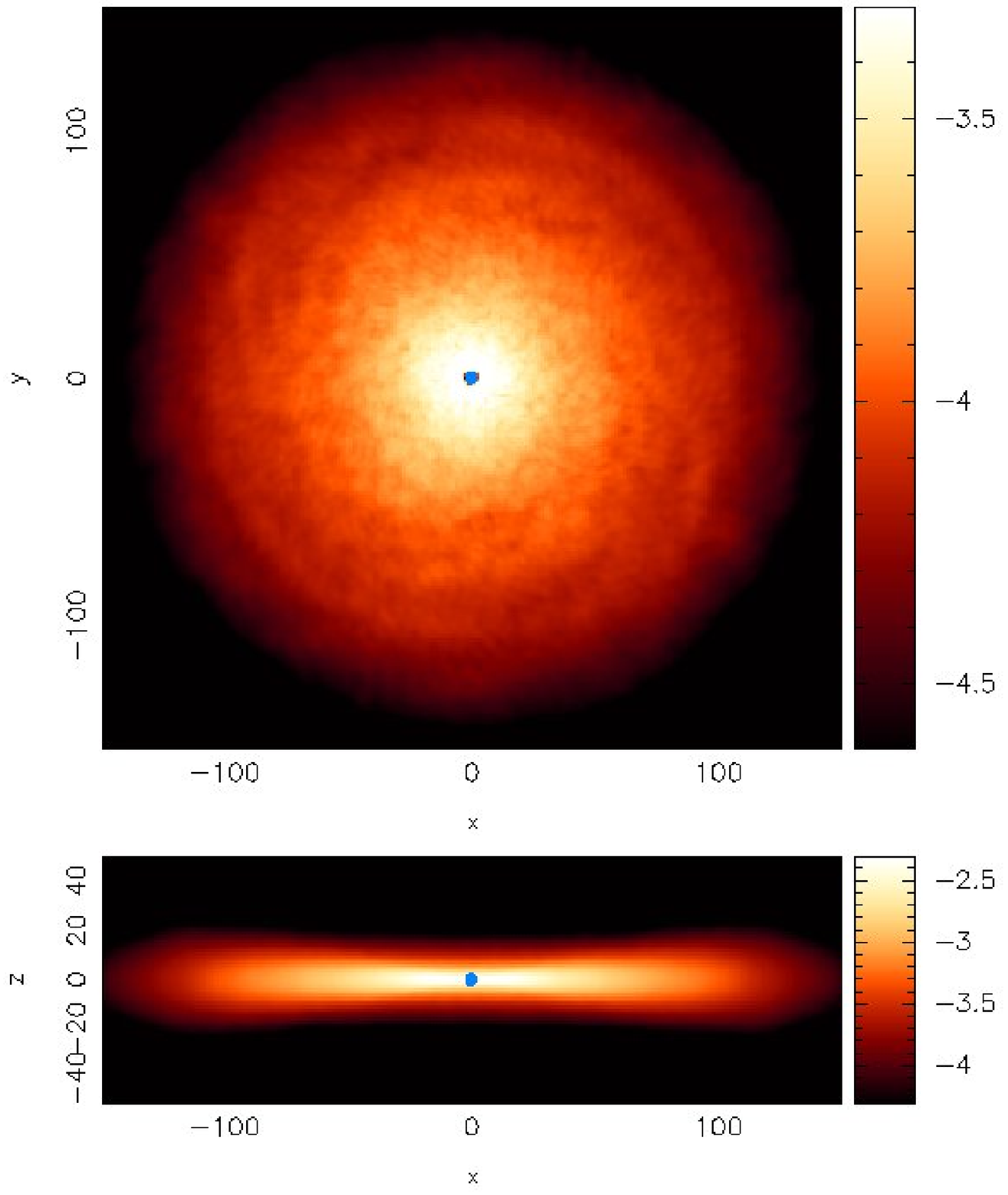}
  \caption{Initial disk conditions.  \textit{Left}: Azimuthally averaged temperature, surface density, and Toomre Q profiles.  \textit{Right}: Face-on and edge-on log-surface density plots of the disk.  Axes are in AU, and surface density is in \msuns AU$^{-2}$.}
  \label{initial_conds}
\end{figure*}
The artificial viscosity formulation used in GADGET-2 is a variation of the standard Monaghan-Balsara viscosity \citep{mon83,bal95}, suggested by \citet{mon97}, utilizing a signal velocity between two particles.  This is implemented in GADGET-2 by
\begin{equation}
  v_{ij}^{sig} = c_i + c_j - 3w_{ij},
\end{equation}
where $c$ is the particle's sound speed.  The viscous tensor is then
\begin{equation}
  \Pi_{ij} = -\frac{\alpha}{2} \frac{v_{ij}^{sig} w}{\rho_{ij}}.
\end{equation}
Here $w = {\bf v}_{ij} \cdot {\bf r}_{ij} / |{\bf r}_{ij}|$ for approaching particles, and $w = 0$ for diverging particles.  Subscript $ij$ denotes the arithmetic mean of the quantity for particles $i$ and $j$, and $\alpha$ is the dimensionless viscous parameter.

We have added one further modification to the viscosity, following \citet{mor97}.  Each particle has an associated $\alpha$, evolving between the minimum $\alpha_{-}$ and maximum $\alpha_{+}$ via
\begin{equation}
  \frac{d \alpha}{d t} = -\frac{\alpha - \alpha_{-}}{\tau} + S.
\end{equation}
Viscosity is thus kept at $\alpha_{-}$ unless the source term $S$ acts upon it, after which it decays with a characteristic time $\tau = h/(kc)$, where $k$ is a dimensionless parameter and $h$ is the SPH smoothing length.  The source term, meant to detect the presence of shocks, takes the form
\begin{equation}
  S = f \alpha_{+} max(-|{\bf \nabla} \cdot {\bf v}|,0),
\end{equation}
with $\alpha_{+}$ our maximum allowed value of the viscous parameter and $f$ a shear viscosity limiter \citep{bal95}.  We use the parameters $\alpha_{-} = 0.1$, $\alpha_{+} = 0.8$ (typical values for $\alpha$ in the unmodified viscosity implemented in GADGET-2 are $\sim$ 0.75), and $k = 2.5$.  Note that GADGET-2 defines the smoothing region over one smoothing length, so that this is equivalent to $k = 5$ for most other SPH codes.  A similar modification to the viscosity in GADGET-2 is used by \citet{dol05}.

\section{INITIAL CONDITIONS}
In order to study interactions between a star-disk system and an impactor, we first need a stable disk around the primary.  We set up the disk with $\sim 4.2 \times 10^{4}$ SPH particles, using the surface density profile 
\begin{equation}
  \label{sigma}
  \Sigma(r) = \Sigma_{0} \left( \frac{r}{r_{0}} \right)^{-1}.
\end{equation}
 
 The vertical density structure at a given radius is given by
\begin{equation}
  \label{rho_init}
  \rho(r,z) = \rho_0(r) exp\left( -\frac{z^2}{2H(r)^2} \right),
\end{equation}
where $H(r) \sim r [c_{s}(r) / v_{\phi}(r)]$ is a temperature dependent scale-height and $\rho_0$ the density at the mid-plane.  The number of particles used is sufficient that the average smoothing length satisfies $h < H(r)/2$ for all but the innermost particles.  The particles are initially placed in pairs, symmetric about the central star, at positions randomized slightly around a grid which would give the distribution of equations (\ref{sigma}) and (\ref{rho_init}).  The simulation is run with just the primary star and the disk until the particle distribution reaches an equilibrium configuration.
 
\begin{figure*}
 \centering
  \plottwo{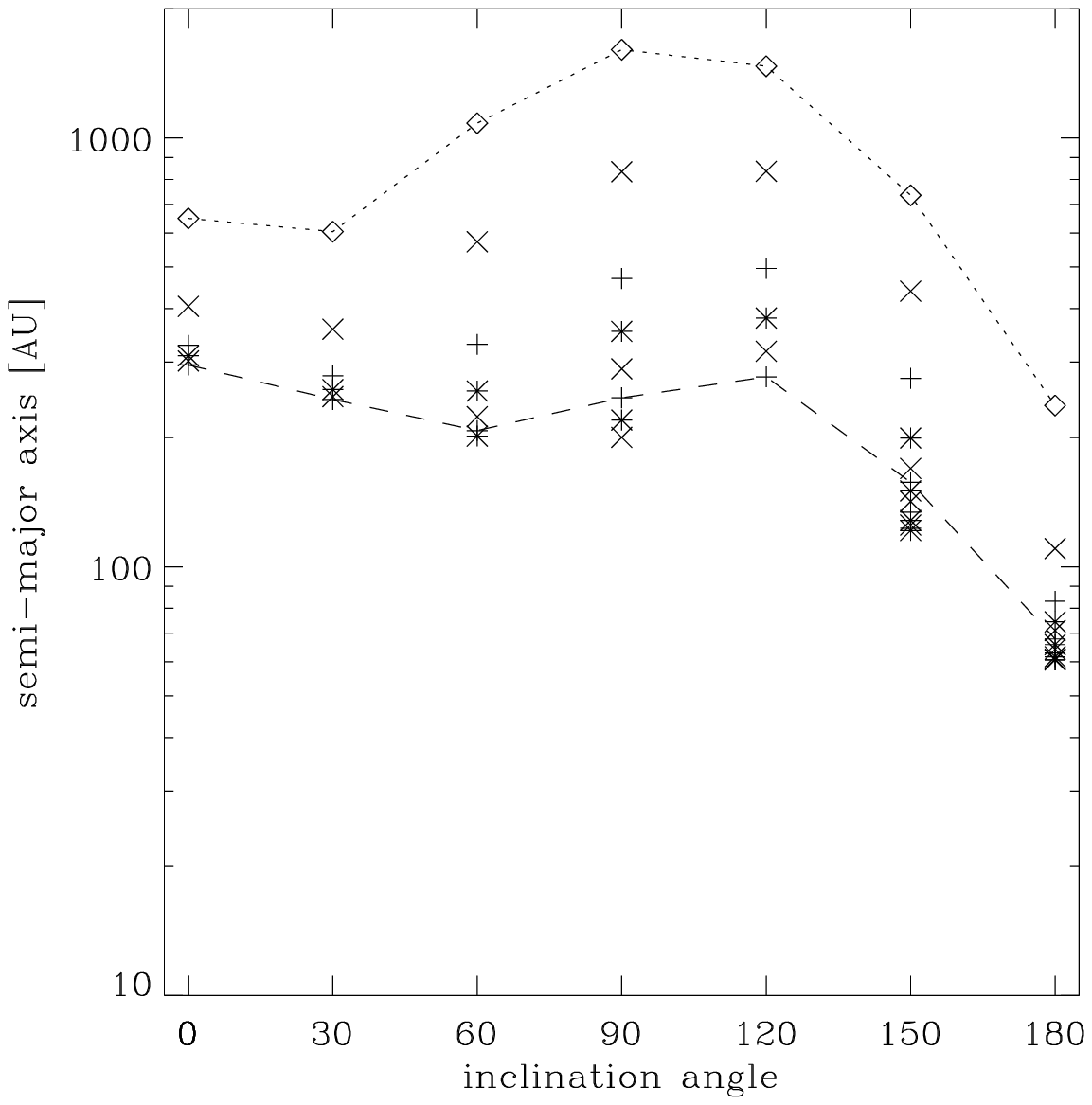}{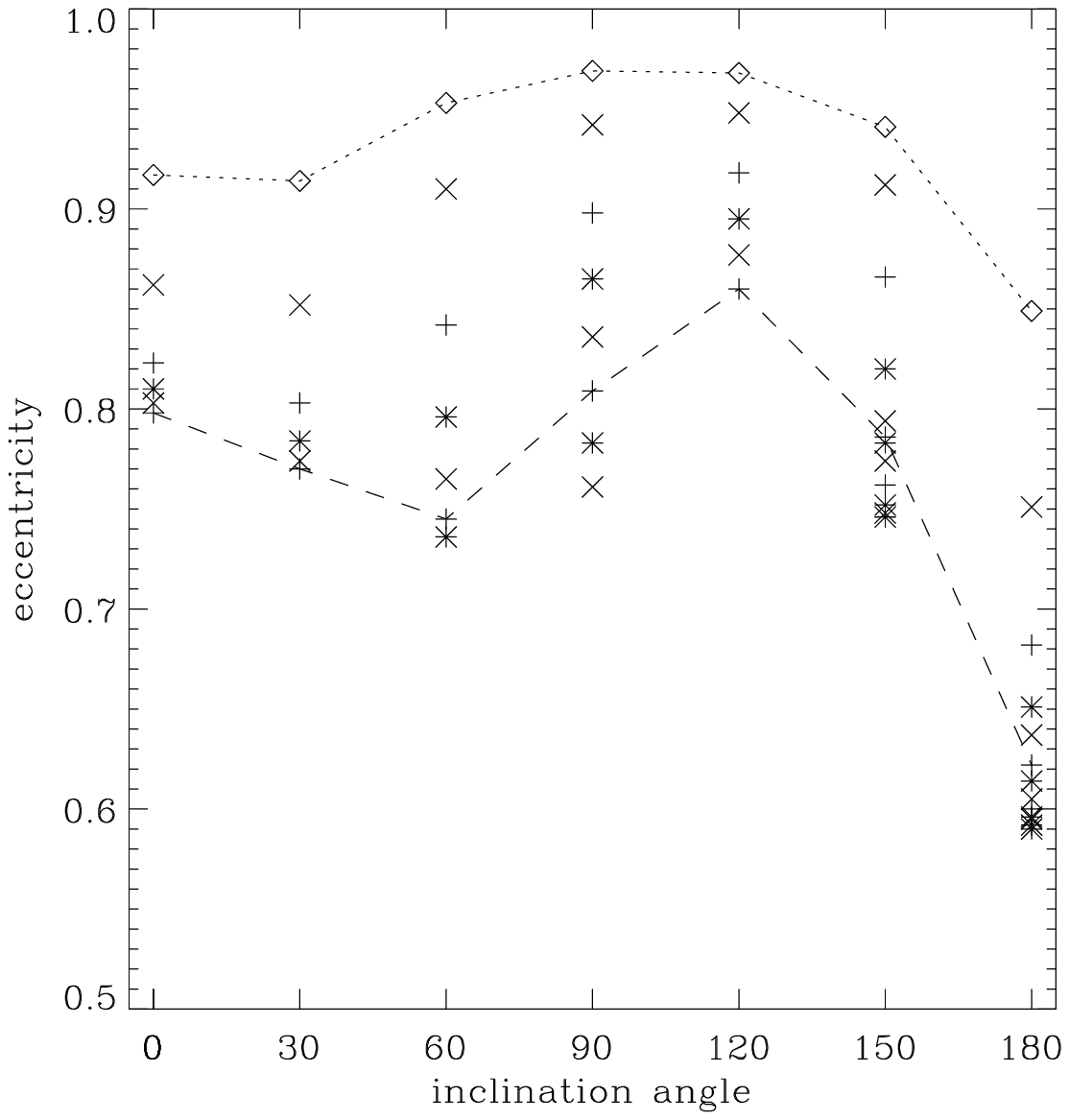}
  \caption{Semi-major axis and eccentricity evolution for the various inclination angles.  The top symbols are after the first encounter, followed by the values after subsequent even numbered encounters.  The dotted line connects the values after 1 encounter, and the dashed line connects the values after 10 encounters.}
  \label{inclination_results1}
\end{figure*}

We use a temperature profile
\begin{equation}
  \label{temp_profile}
  T(r) = T_{0} \left( \frac{r}{r_{0}} \right)^{-1/2},
\end{equation}
where T$_{0}$ is set depending on the mass of the star the particle is closest to.  We implement a simple cooling scheme to maintain this temperature profile while the disk is unperturbed,
\begin{equation}
  \label{cooling}
  \frac{d u}{d t} = -\frac{u - u_{base}}{\tau_c}.
\end{equation}
The timescale $\tau_c$ is inversely proportional to the Keplerian orbital frequency at the particle's radius, and $u_{base}$ is the internal energy corresponding to the temperature profile given by equation (\ref{temp_profile}).  The cooling time we use is short enough that the disk would eventually fragment \citep{gam01,ric05} in the absence of the base temperature profile, below which the particles are not allowed to cool.  Since we are interested in disks that are stable for long time periods, we choose initial conditions such that the Toomre parameter Q is greater than unity at all radii.  Figure \ref{initial_conds} shows the temperature, density and Toomre Q profiles of the equilibrium disk, as well as column density plots.

\section{ANALYTICALLY ADVANCING SIMULATIONS} \label{analyticadvance}
In order to save computational time, if an encounter results in a binary with a semi-major axis much larger than the disk radius the system is not integrated through apastron and back.  Instead, once a simulation has advanced so that the stars are far enough apart that tidal effects on the remnant disk are no longer important, and the binary's semi-major axis and eccentricity stop evolving, numerical integration is stopped.  This separation is typically several initial disk radii.  By analytically advancing the system along a Keplerian orbit until the effects of the disk again become important, we can reduce the computation time by up to an order of magnitude for several of the cases considered.   

When a system is ready to be analytically advanced, the particles are grouped into one of three populations; those which have been ejected, those that are bound to the impactor, and those bound to the primary.  If a particle's total energy relative to the stars' center of mass is positive, it is ejected and removed from the calculation.  The remaining particles are assigned to a star based on which potential well is deeper at that location.  Though simplistic, this scheme gives good results once the stars separate enough that two clear populations are visible.  At this point we treat the populations as point masses at their centers of mass, and determine an elliptical orbit.  The two stars and their associated gas particles are advanced analytically along this orbit until the point when the separation is 350 AU. 

Upon restarting, the particles in each group are placed using a circularizing scheme similar to that employed by \citet{hal97}, as follows.  The angular momentum of the particle about the star is determined, as well as the separation vector from the star to the particle.  The separation vector is stretched to the distance of a Keplerian orbit with the same angular momentum.  The velocity is made Keplerian, with the same orientation of the angular momentum vector, and the particle is advanced along its new orbit for the length of time between stopping the simulations and restarting it.  

 This circularizing scheme is an approximation, and one that is probably not accurate for material which is on the verge of being unbound, and is now on a highly eccentric orbit.  However, most of the material is in a perturbed disk close to the primary star, and circularizes fairly quickly if the simulation is allowed to continue.  For this material, the procedure is reasonable as long as the orbital period of the binary is long compared to the orbital period of the disk particles.  The viscous time scale for $r > 20$ AU is greater than the longest binary orbital period, so we can ignore viscous spreading of the disk between encounters.  This technique is only applied to simulations with times between encounters $\sim 10^{3}$ yr; when the system evolves to the point where the time between periastra is $\sim 10^{2}$ yr, comparable to the orbital timescale at the edge of the disk, numerical integration is allowed to continue uninterrupted.

\section{RESULTS}
In this study, we keep the mass of the primary star fixed at $m_{p} = 20$ \msuns, and the disk mass at $m_{d} = 2$ \msun.  The main group of simulations are between that star-disk system and a $m_{i} = 5$ \msuns impactor.  The impactor is introduced on a slightly hyperbolic ($v_{\infty} = 0.5$ \kms) orbit that, unaltered by the disk, would have a periastron distance of 50 AU.  The initial separation between stars is 350 AU.  This group of simulations covers the range of inclination angles $i = $[0,180] degrees, in steps of 30 degrees, and each set of initial conditions is advanced through at least 10 encounters.  The retrograde, coplanar case ($i = 180$) was also run with an impactor mass of 1 \msuns and a 50 AU periastron, and with a 5 \msuns impactor at a periastron of 100 AU.  For simulations which are analytically advanced, the orbital characteristics discussed below are measured when the simulation is stopped; for those allowed to run continuously, they are measured at the binary apastron.

\subsection{Inclination Angle}

We introduce the 5 \msuns impactor such that the angular momentum vector of the disk lies in the plane defined by the angular momentum of the orbit and the semi-major axis of the orbit, so that out of plane encounters pass through the disk twice.  An inclination of $i = 0$, a prograde encounter, is when the disk and orbital angular momenta are aligned; a retrograde encounter, $i = 180$, has the angular momenta anti-aligned.  All of the simulations considered here are penetrating encounters, similar to SPH simulations by \citet{hel95} and \citet{bof98}, and reduced three-body calculations by \citet{hal96},  though at a much higher mass range, and different mass ratios.  Like those authors, we find that the dissipation of orbital energy is most efficient for retrograde encounters.  

\begin{figure}
 \centering
  \plotone{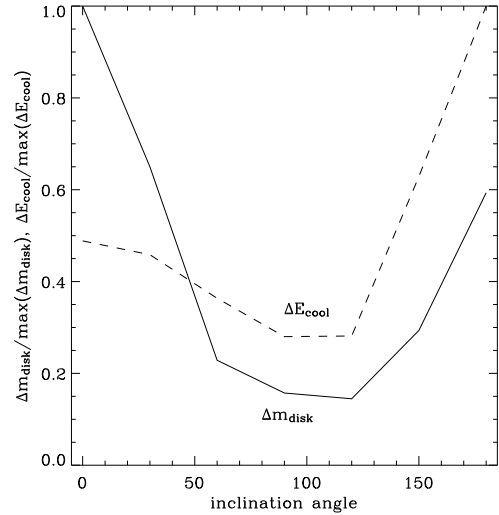}
  \caption{Disk disruption (\textit{solid}) and energy loss from cooling (\textit{dash}), relative to the maximum value for all inclinations.  Disk disruption is a more important energy transfer mechanism for low inclinations, while losses from shock heating (measured by the subsequent cooling) dominate for retrograde encounters.}
  \label{dissipation_cooling}
\end{figure}

Figure \ref{inclination_results1} shows the long term (over many binary orbits) evolution of the binary semi-major axis and eccentricity for each inclination considered.  The dotted line connects the values after one encounter, and the dashed line connects the values after 10 binary orbits. 
The semi-major axis $a$ is related to the specific orbital energy $E$ and the total binary mass $M = m_{p} + m_{i}$ by $a = GM/2|E|$.  For the purposes of determining orbital characteristics, the gas mass is distributed between the primary and the impactor as described in section \ref{analyticadvance}.  Plotted in figure \ref{dissipation_cooling} is a rough measure of the importance of disk disruption and heating by shocks in the dissipation of orbital energy.  Disk disruption is measured by $\Delta(m_{p} + m_{d})$, while the energy lost due to shock heating of the disk gas is measured by the total cooling energy (see eq. [\ref{cooling}]) calculated during the encounter.  Disk disruption is maximal for a prograde encounter, while heating losses are maximal for retrograde encounters.  

Disk disruption is not a reliable tracer of energy transfer; \citet{hal96} note that while a prograde encounter is the most spatially destructive, a retrograde encounter has a larger energy transfer.  Nonetheless it provides some idea about the mechanism behind the energy loss from the orbit.  An encounter with $i =150$ has $\sim 1.4$ times the energy loss from heating as a prograde or $i = 30$ encounter, but a larger semi-major axis after the collision.  The dominant energy transfer mechanism for these nearly-prograde encounters is evidently not the shock heating of disk material but the gravitational interactions with, and disruption of, the initial disk, evidenced by over twice as much disk material being lost from the primary for $i \leqq 30$ than for $i = 150$.  This difference in dissipation mechanism is noted by \citet{hel95} and \citet{bof98}, and the general shape of the semi-major axis plot in figure \ref{inclination_results1} is consistent with those author's findings for $\Delta E_{orbit}$ with inclination angle.
\begin{equation}
  \label{periastra}
  r_{peri} = a (1 - e),
\end{equation}
with the eccentricity given by 
\begin{equation}
  \label{eccentricity}
  e = \sqrt{1 + \frac{2EJ^{2}}{G^{2}M^{2}}}.
\end{equation}
We run all simulations until either the semi-major axis is essentially unchanged with further orbits, or the disk is destroyed, with two exceptions.  The 90 and 120 degree inclination cases continue to evolve after 14 and 11 orbits respectively, and their disks retain $\sim 1$ \msun.  However, the elapsed times for these simulations are approaching $3 \times 10^4$ yr.  Note that the retrograde case has undergone 27 orbits in less than $4 \times 10^3$ yr.  For the environments we are considering, dense cluster cores, the timescale and size of the orbits ($a \sim 1500$ AU) for these cases are too large to ignore perturbations from other cluster members or the addition of mass from the cloud core.
\begin{figure*}
 \centering
  \plottwo{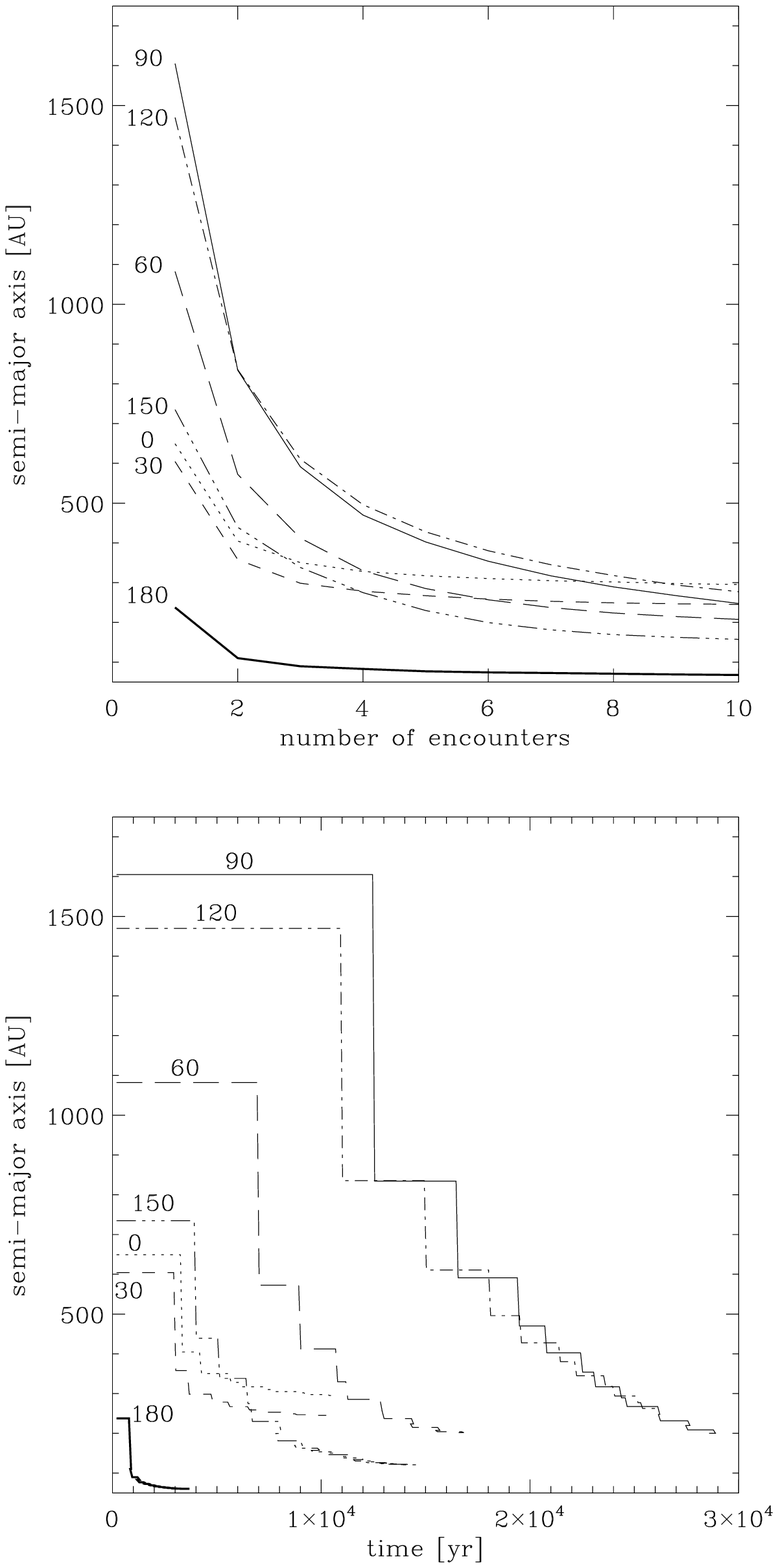}{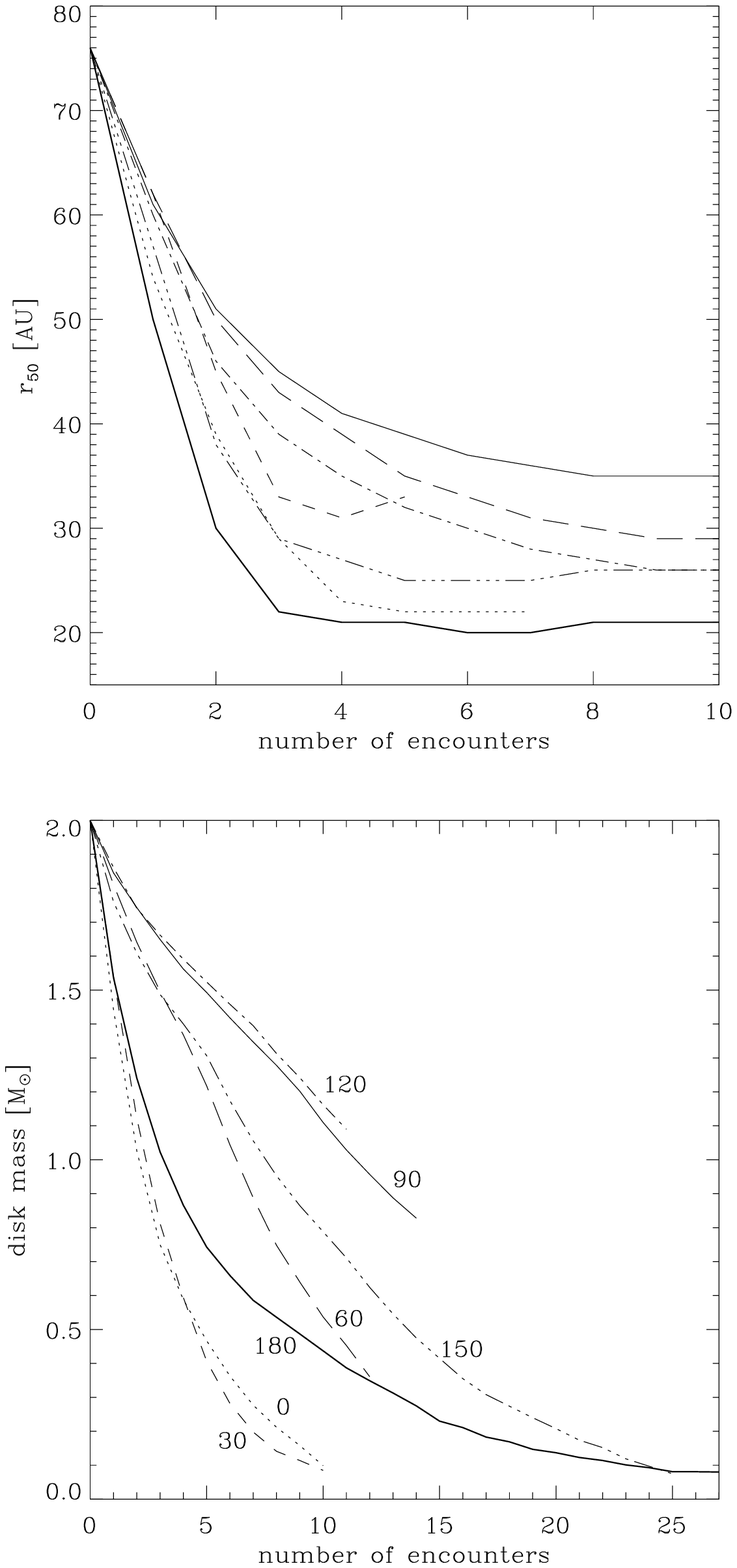}
  \caption{\textit{Left}: Semi-major axis evolution for the various inclination angles, as a function of (\textit{top}) encounter number and (\textit{bottom}) time since the first encounter.  For clarity, values in time are plotted as step functions measured at apastron.  \textit{Top right}: Disk half-mass radius as a function of encounter number.  \textit{Bottom right}: Remaining disk mass as a function of encounter number.  Values are measured at apastron, or when the simulation is stopped for analytic advancing.}
  \label{inclination_results2}
\end{figure*}
Plotted in figure \ref{inclination_results2} are the semi-major axes as functions of encounter number and time, and the half-mass radii and masses of the disks after each encounter.  The disk's half-mass radius is defined as the radius inside of which 50\% of the gas energetically assigned to the primary is located.  The general trend for all inclinations is a period lasting several encounters during which a steady decrease in orbital energy takes place.  After a few encounters, this rate decreases and the change in orbital parameters levels off.  Note that the prograde case, while moderately effective at dissipating orbital energy over a few encounters, has the largest semi-major axis after 10 encounters.  The reason for this is a combination of two effects.  The high rate of disk disruption for low inclinations is a major factor; the prograde case has reduced the disk mass from 2.0 to 0.5 \msuns after only 5 orbits, whereas the retrograde case takes 10 orbits.  This work and previous work \citep{hel95,hal96,bof98,pfa05} verify the intuitive idea that with a larger disk mass, the orbital energy decay is greater.

\begin{figure}
 \centering
  \plotone{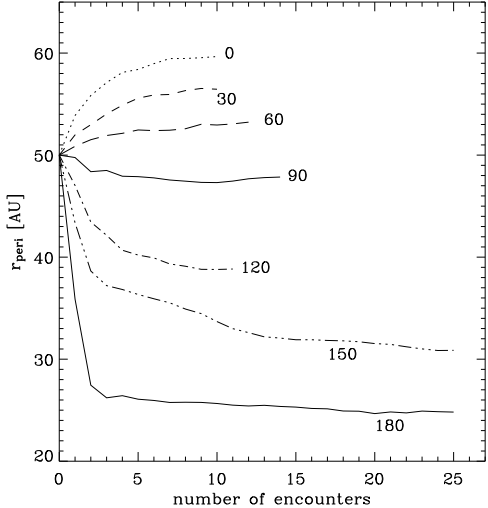}
  \caption{Periastron separation for each inclination as a function of encounter number.}
  \label{periastron_evolution}
\end{figure}

The evolution of the periastron distance also helps encounters with a retrograde component achieve faster binary decay during the disk destruction phase.  Plotted in figure \ref{periastron_evolution} are the periastra after each encounter for all inclinations.  The periastra can be calculated from the orbital parameters by 

Thus, the periastron distance is a function of the orbital energy and angular momentum.  Note that $E$ is negative for an elliptical orbit; for a given decrease in $E$, there is a critical change in $J$, above which the periastron will increase.  Encounters with $i < 90$ don't lose enough orbital angular momentum (through accretion and gravitational interactions) to decrease the periastron distance, so while the disk radius is decreasing, the encounter distance increases.  This behavior is due in part to the location of the corotation resonance region for the different prograde inclinations \citep{hal96}.  Material in the resonance region tends to transfer angular momentum from the disk to the binary \citep{ost94}, working against the dissipative effects of accretion and material outside the corotation radius.

Previous work on similar encounters \citep{hel95,hal96,bof98,pfa05} shows that with increasing periastron distance, the magnitude of the change in orbital energy decreases.  Retrograde encounters, however, show a large decrease in periastra, down to approximately the disk half mass radius.  Comparing the disk characteristics for $i = 0$ and $i = 180$ after 5 encounters (Figure \ref{inclination_results1}) with their periastra evolution, we see that while $r_{50} \sim 22$ AU for both cases, the closest approach for $i = 180$ is $\sim 26$ AU, still close to the disk, while for $i = 0$ the closest approach is nearly 60 AU, so that $r_{peri} \gtrsim 2 r_{50}$.

\begin{figure*}
 \centering
  \plotone{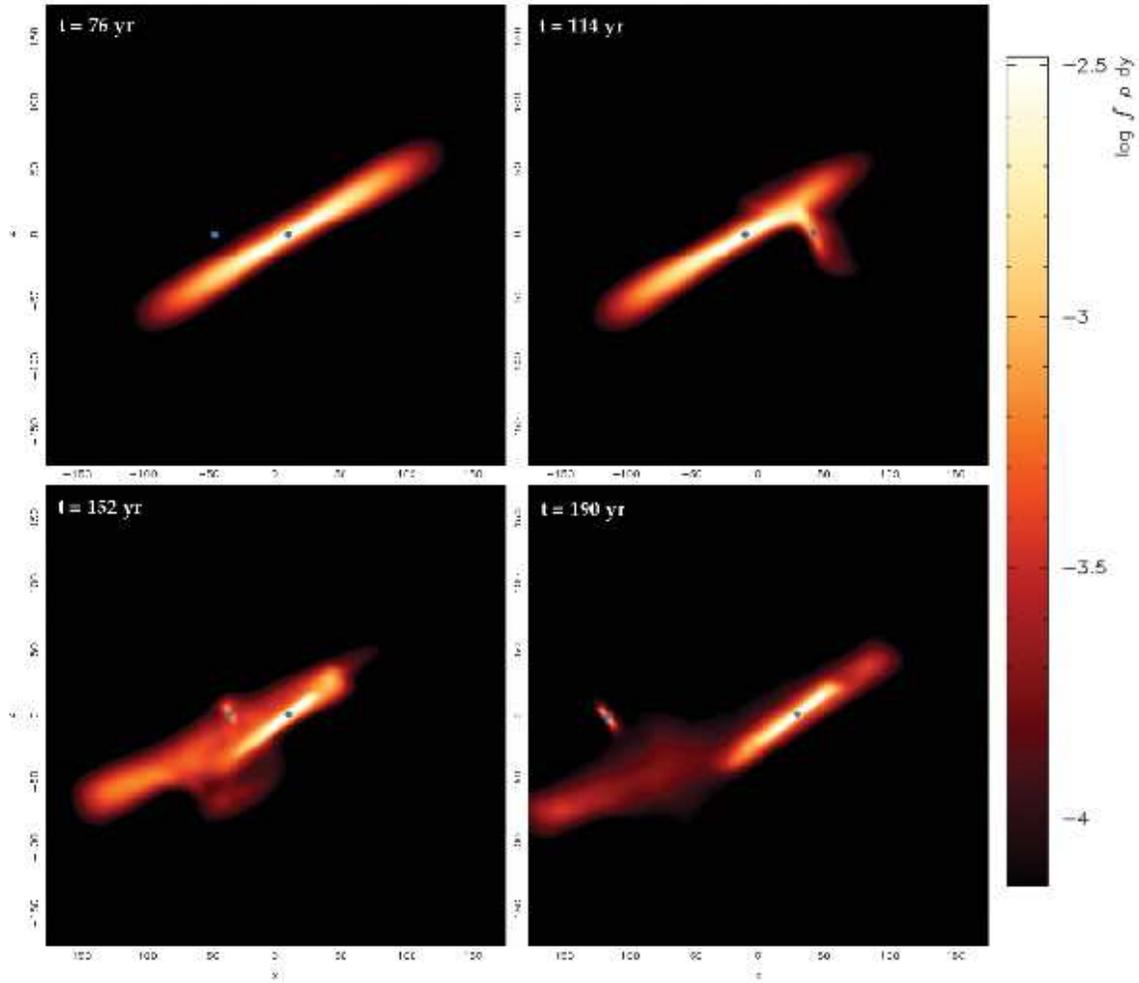}
  \caption{Edge-on snapshots of the first encounter for $i = 30$.  Plotted is log column density, in units of \msuns AU$^{-2}$}
  \label{30edgecombo}
\end{figure*}

In several cases with a prograde encounter, mass is transfered from the disk, resulting in a secondary disk around the impactor.  This is illustrated in figure \ref{30edgecombo} for $i = 30$, the most striking example.  In this case the disk is nearly orthogonal to the primary disk, a scenario tantalizingly similar to the non-coplanar disks seen in young multiple systems such as HH111 and HH121 \citep{rei99}.  The disk survives (or is replenished) during subsequent encounters.  Increasing periastron separation and rapid disk destruction are seen in both the prograde encounter and for $i = 30$; we suspect that the increased interaction between the disks causes the out-of-plane encounter to end up in a tighter binary.

\begin{figure}
 \centering
  \plotone{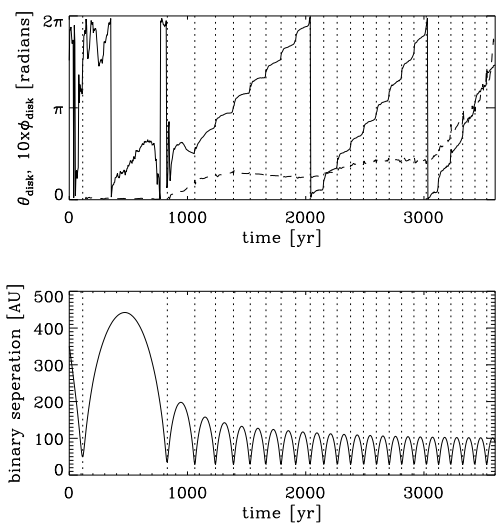}
  \caption{\textit{Top}: Inner disk orientation ($r < 30$ AU) as a function of time, for i = 180.  Plotted are the azimuthal angle $\theta$ (solid) and ten times the polar angle $\phi$ (dashed) relative to the initial disk orientation.  \textit{Bottom}: Binary separation for the same simulation.  Periastra are marked by dotted lines.}
  \label{disk_orientation}
\end{figure}

\begin{figure}
 \centering
  \plotone{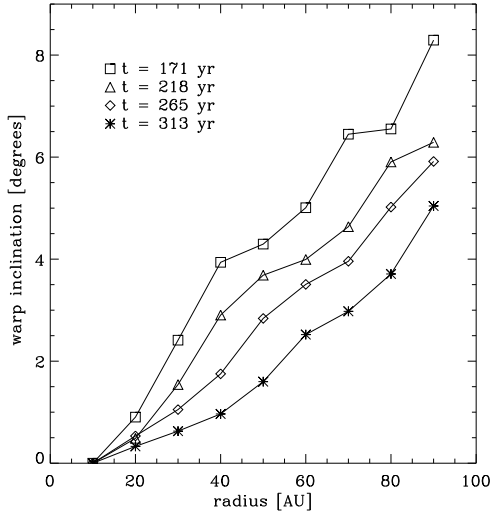}
  \caption{Warping of the 30 degree inclination disk, measured as the difference in orientation between annuli 10 AU wide and the central annulus.  Times are measured from the first periastron.}
  \label{warping}
\end{figure}

\begin{figure}
 \centering
  \plotone{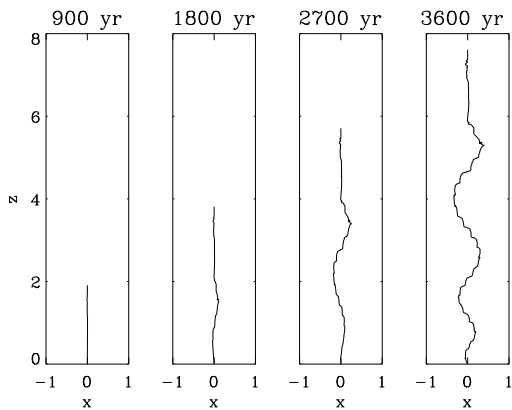}
  \caption{\textit{Top}: A toy jet launched from the same disk as figure \ref{disk_orientation}.  The jet is assumed to launch perpendicular to the inner disk, and with a uniform velocity throughout the evolution.}
  \label{jet_snapshots}
\end{figure}

An effect noted in previous work \citep{hel95,lar97,bof98} is that out-of-plane encounters lead to tilting of the remnant disk.  After several encounters, this effect is seen even in the in-plane, retrograde case.  When followed through several orbits, this effect manifests itself as stepped precession, shown in figure \ref{disk_orientation}.  The orientation of the disk is measured as the average angular momentum of gas particles about the primary star.  In figure \ref{disk_orientation} we consider only the inner disk, $r < 30$ AU.  The disk undergoes a slow precession during most of the orbit, with periodic jumps associated with periastron passages.  There is warping present in the disks, shown in figure \ref{warping} for the 30 degree inclination case.  Plotted is the difference in the orientation of annuli 10 AU wide at four different times, relative to the central annulus.  This angle is measured by $i = cos^{-1}[{\bf L}_{a} \cdot {\bf L}_{b} / (|{\bf L}_{a}||{\bf L}_{b}|)]$, with ${\bf L}_{a}$ and ${\bf L}_{b}$ the summed angular momenta of all particles in two annuli.  If the orientation of an outflow is tied to the inner disk angular momentum vector, this is potentially an observable effect \citep{bat00a}.  
Shown in figure \ref{jet_snapshots} is a crude estimation of what such an outflow might look like, again for the retrograde case.  We assume that a jet is launched along the disk angular momentum vector, at a uniform velocity for the entirety of the simulation.  The units are arbitrary and depend on the velocity of the outflow; for a 10 \kmss jet, the long axis of the jet extends $\sim 8000$ AU.  It isn't until $\sim 2000$ years, or 10 binary orbits, that the jet begins to precess noticeably.  As the encounters become more frequent and the disk becomes slightly non-coplanar to the binary orbit, the precession becomes more pronounced.  The sharp kinks in the jet are a result of the torque being applied almost totally during periastra passages.

\subsection{Evolution of a Binary in an Accreting Envelope}
Material from the parent cloud that accretes onto a binary can alter the binary orbital elements and mass ratio, effects that have been studied by several authors \citep{art83,bat97,bat97b,bat00}.  \citet{art83} suggests that low angular momentum gas falling nearly radially onto the system will result in a decrease of the semi-major axis, a result confirmed by \citet{bat97b}.  \citet{bat00} shows that the decrease in separation can reach an order of magnitude or more, provided that the mass accreted onto the system is several to 10 times the total stellar mass.  \citet{bat02} describe a system whose components increase their masses several fold, while the separation decreases from $\sim 100$ to $\sim 10$ AU.  \citet{bon05} show that accretion onto a low mass ($m \sim 1$ \msun) binary with an initial separation $\sim 1000$ AU can lead to a close (separations $\sim 1$ AU or less) binary with a total mass $\sim 30-50$ \msun.  In all of these studies the accreted mass is comparable to, or orders of magnitude larger than, the initial system mass. 

In order to see a comparable effect with a total stellar mass of $\sim 25$ \msun, a much larger mass must be added to the system.  If the system mass is to double in less than $10^{5}$ yr, an average accretion rate of $\dot{M} > 2.5 \times 10^{-4}$ \msuns yr$^{-1}$ is required, which is fairly high, though perhaps not unreasonable for the formation site of a 20 \msuns star \citep{mck03}.  Given the previous results, even this high rate would result in a hard binary only after $\sim 10^{5}$ yr, which is long compared to the encounter time.  It seems that in massive systems such as those considered here, accretion from a surrounding envelope will not be a large factor in hardening binaries.  However, the previous work has been mostly concerned with less massive, and less eccentric binaries.  In order to explore this issue, we consider the effect of a massive accreting envelope in the retrograde case.

We add an in-falling envelope with $\dot{M} = 10^{-3}$ \msuns yr$^{-1}$, consisting of $\sim 5 \times 10^{4}$ particles, originating at a radius of 500 AU from the center of mass of the system.  The in-fall is spherically symmetric, purely radial, and is introduced with a velocity equal to the free fall velocity at 500 AU.  The lack of rotation is unrealistic, but chosen for maximal effect on the orbit.  The cutoff radius of 500 AU was chosen to keep the simulation time reasonable.  The effect of any ejected mass on shaping the in-fall at larger radii is thus effectively negated.  A bigger concern is the lack of radiation, which can effect in-fall morphology significantly \citep{yor99,yor02,kru05b}.  Spherical in-fall, and thus the lack of angular momentum in the envelope, should lead to a maximal decrease in semi-major axis, so this simulation is illustrative of the most extreme case.

\begin{figure*}
 \centering
  \plottwo{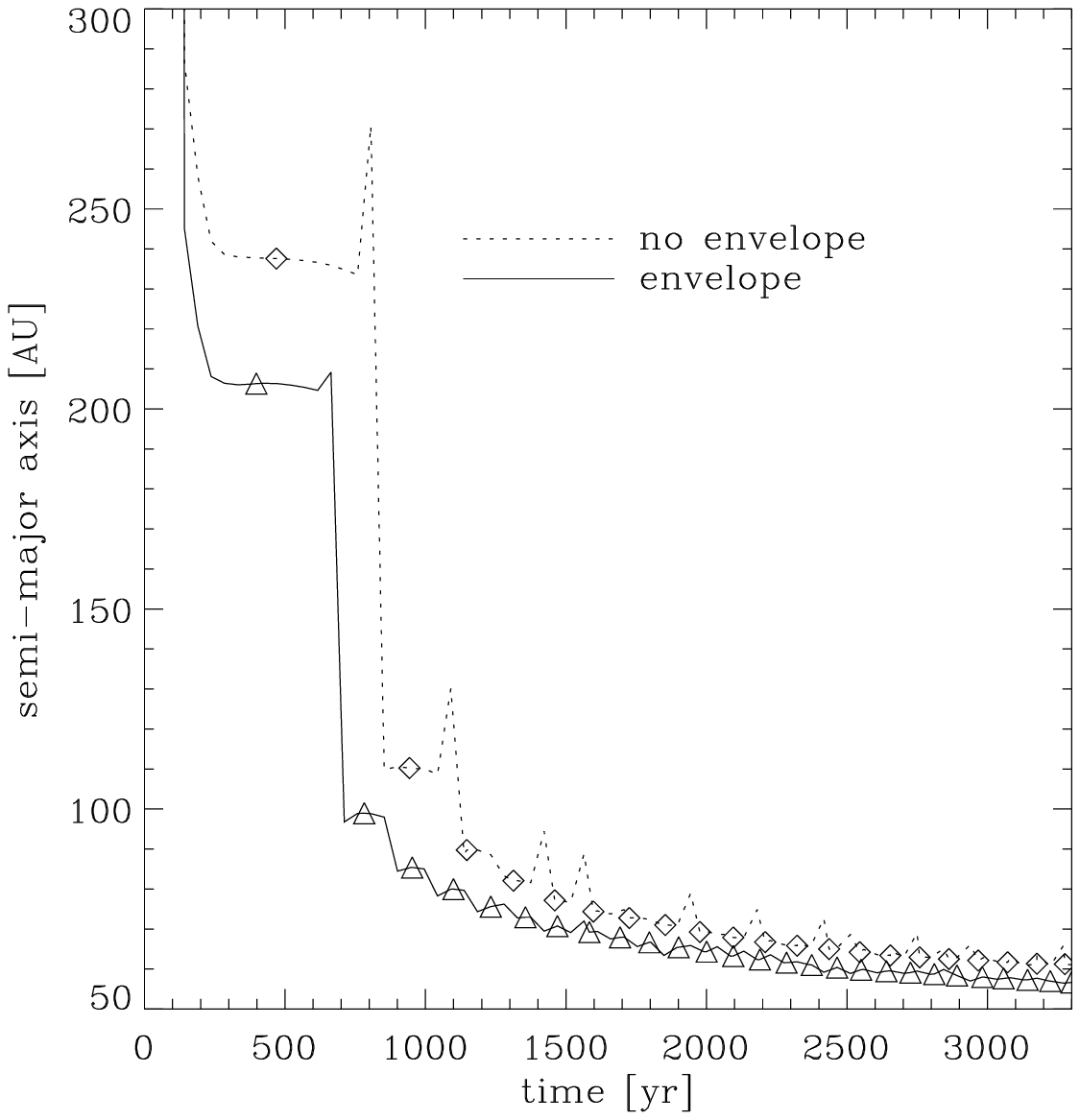}{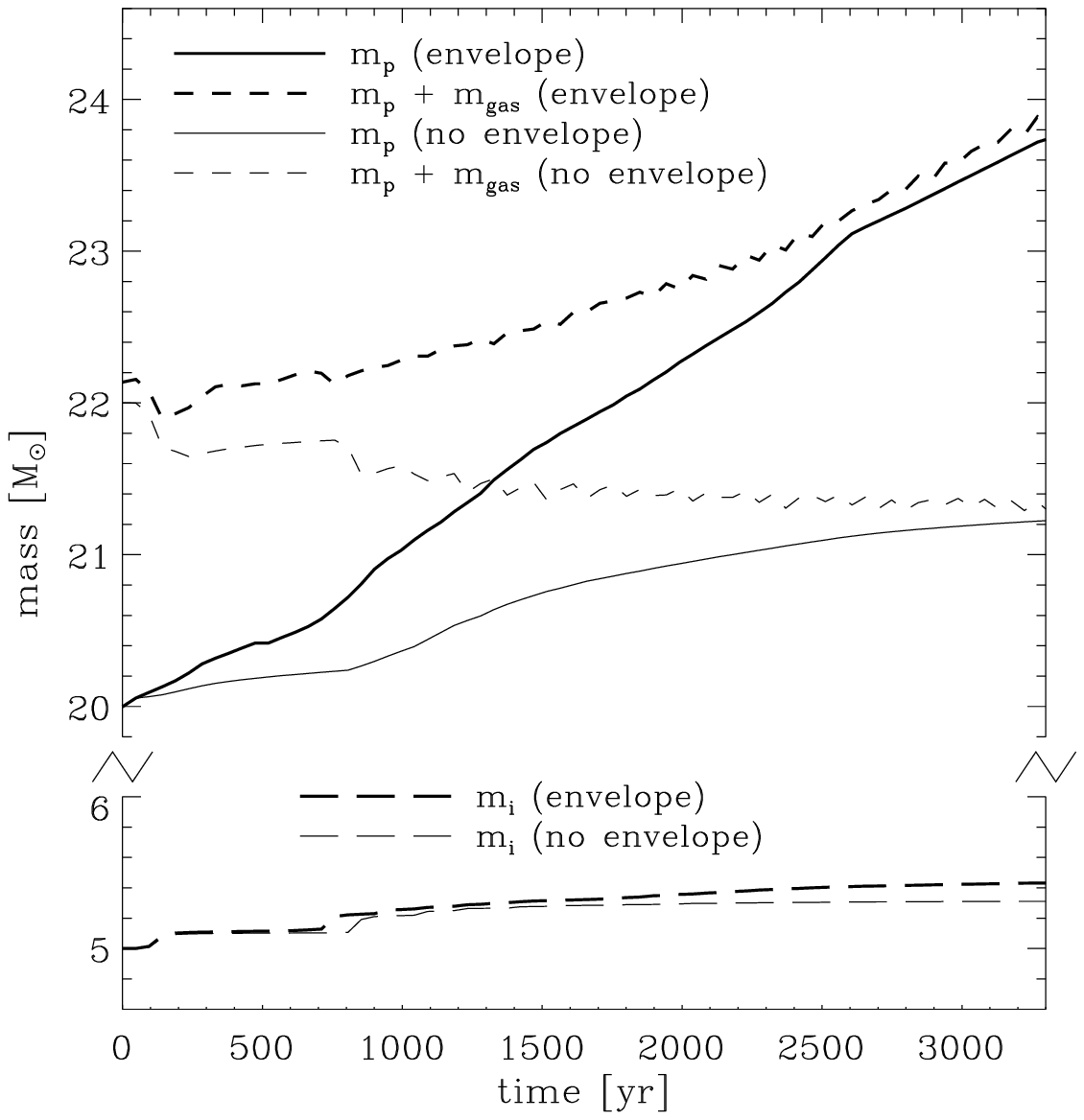}
  \caption{\textit{Left}: Semi-major axis evolution with time for i=180, with (\textit{solid}) and without (\textit{dot}) an accreting envelope. Apastra are marked with symbols.  \textit{Right}: Impactor and primary masses, and mass of gas associated with the primary for the two cases.}
  \label{envelope_results}
\end{figure*}

Plotted in figure \ref{envelope_results} are the semi-major axis and mass evolution of the retrograde case with and without an in-falling envelope.  Spikes in the semi-major axis plot are an artifact of our orbital parameter calculation scheme during periastra approaches.  The gas mass plotted is that which is energetically bound to the primary star.  The initial binary decay is faster for the case with the in-falling envelope, though the final state of the system is similar to the case with no envelope.  Comparing the masses of the stars for the two cases, it is apparent that most of the infalling mass is accreting onto the central star.  At $t \sim 3000$ yr the impactor in the envelope case has a mass $m_{i} \sim 5.4$ \msun, only slightly more than $m_{i} \sim$ 5.3 \msuns in the absence of an envelope.

At $t \sim 3000$ the primary mass without an envelope has leveled off at $m_{p} \sim 21.1$ \msun, and the disk mass (the difference between $m_{p}$ and $m_{p} + m_{gas}$) has decreased to $m_{d} \sim 0.1$ \msun.  With the addition of a spherical infall, the disk mass is minimally effected, again approaching negligible mass after $t \sim 2500$ yr, at which point the primary star enters a period of steady growth at $\dot{M} \sim 10^{-3}$ \msuns yr$^{-1}$.  The primary itself has grown to $m_{p} > 23$ \msun; it appears that the infall is being funneled almost entirely onto the primary, spending little time in the disk.  Because we were searching for a purely accretional effect as opposed to the influence of a circumbinary disk \citep{bat97b}, the infall was set up with no angular momentum.  In this case, once the system has evolved to a near steady state, with grazing disk encounters slowly eroding the binary energy and circularizing the orbit, the stellar masses dominate the mass accreted from the envelope and the system is minimally affected, as expected.

The total mass accreted from the envelope over 3000 yr, while small compared to the stellar masses, is greater than that of the initial disk; if the angular momentum of the envelope is sufficient to accrete into a disk around the primary at a radius comparable to the binary orbit, the initial period of rapid orbital decay may be extended and tighter binaries could be formed.  However, accretion of low angular momentum gas from the parent core seems unlikely to significantly alter the orbits of the massive capture formed binaries studied here.    
  
\section{SUMMARY AND CONCLUSIONS}
We have presented a suite of simulations which consider the dissipative effects of a circumstellar disk on a capture formed binary, in the context of a massive protostar encountering a less massive cluster member.  The systems were followed for 10 or more orbits, in many cases leading to almost total destruction of the original disk.  We find that for the parameters studied, the interactions result in a soft, eccentric binary system.  The energy change associated with different inclination angles is in agreement with previous simulation results.  The periodic torquing of the disk by the passage of the impactor may be observable in the orientation of a jet or outflow.

These simulations apply to a massive protostar that has accreted most of its final mass.  The disk and envelope mass is $\sim$ 10 \% of the stellar mass, analogous to the class I phase of a low mass protostar's evolution.  Previous studies of stellar encounters with star-disk systems \citep{hel95, bof98} concluded that disk capture rates were too low to contribute significantly to the observed multiplicity of low mass stars.  All but the least dissipative encounters simulated here result in a binary with a semi-major axis $\lesssim$ 500 AU within $10^{4}$ yr.  Extending the analysis of binary capture rates to the regime where the disk mass and impactor masses are comparable, and much less than the primary, may show that the high multiplicity of massive stars such as the Trapezium stars in Orion could result from disk encounters in the late stages of the cluster's formation. 

In order to follow the encounters through several orbits, the parameter space covered here is small; only a single impactor mass and initial periastron were considered.  The mass of the impactor was chosen as the minimum such that a merger would lead to significant growth.  As the impactor's mass approaches the primary's, disk destruction occurs over fewer orbits and less time is available to harden the binary.  
The apastra of the resultant binaries are on the order of 100s of AU, and the periastra of the most dissipative cases remain above $\sim 25$ AU.  For the parameters studied, it seems that in the absence of further effects, disk dissipation alone cannot lead directly to mergers.   

Accretion onto the system from the parent core is unlikely to lead to significant hardening of the binary over any short timescale, as the stellar masses considered are too high for reasonable mass accretion rates to have a large effect.  Thus if disk mediated collisions are to contribute to the growth of a massive star, they must occur earlier in the cluster's evolution, when the disk and envelope are more massive and perhaps more dissipative.  Even if this is the case, there must be an efficient disk replenishment mechanism in order for more than a single merger to occur.  However, at least two further mechanisms seem likely to come into play.

The disks considered in this work are Toomre stable; if a colder disk is considered, gravitational fragmentation within the disk will change the nature of the interactions, potentially leading to a more rapid decrease in the binary separation as smaller bodies are ejected from the system.  In a similar dynamical effect, a binary with a semi-major axis $\sim 100$ AU presents a large target for interactions with other encounter partners in the dense cluster core.  In this scenario the capture and gradual tightening of the initial binary, modeled here, provides the initial conditions for encounters with other cluster stars.  The dynamics of $n>2$ bodies plus the remnants of a disk are beyond the scope of this work, but are an enticing direction for further study of systems such as these.
    
In summary, it seems that disk interactions in the late phase of protostellar evolution considered here do not provide a direct path to merger.  Further accretion from the natal cloud will not be enough to induce collisions from the resultant binaries.  In order for disks such as those modeled here to play a role in any collisional growth of massive stars, the interactions must involve third bodies acting on the capture formed binary, through either induced fragmentation of the disk or from other cluster members.  Another, yet unmodeled possibility is that the encounters leading to a merger take place during an evolutionary phase more like a class 0 protostar, with the mass of the primary comparable to that of the disk surrounding it.  Encounters with such systems have only been considered with impactors of similar mass to the primary, and not in the limit where the disk mass is greater than the impactor.

We thank the referee, Ian Bonnell, whose comments helped improved the manuscript.  This work was supported by the NASA Astrobiology Institute through a Cooperative Agreement with the University of Colorado, through grant NNA04CC11A.


\begin{thebibliography}{46}


\bibitem[{{Artymowicz}(1983)}]{art83}
{Artymowicz}, P. 1983, Acta Astronomica, 33, 223

\bibitem[{{Bally} \& {Zinnecker}(2005)}]{bal05}
{Bally}, J., \& {Zinnecker}, H. 2005, \aj, 129, 2281

\bibitem[{{Balsara}(1995)}]{bal95}
{Balsara}, D.~S. 1995, Journal of Computational Physics, 121, 357

\bibitem[{{Bate}(1997)}]{bat97}
{Bate}, M.~R. 1997, \mnras, 285, 16

\bibitem[{{Bate}(2000)}]{bat00}
---. 2000, \mnras, 314, 33

\bibitem[{{Bate} \& {Bonnell}(1997)}]{bat97b}
{Bate}, M.~R., \& {Bonnell}, I.~A. 1997, \mnras, 285, 33

\bibitem[{{Bate} {et~al.}(2002){Bate}, {Bonnell}, \& {Bromm}}]{bat02}
{Bate}, M.~R., {Bonnell}, I.~A., \& {Bromm}, V. 2002, \mnras, 336, 705

\bibitem[{{Bate} {et~al.}(2000){Bate}, {Bonnell}, {Clarke}, {Lubow}, {Ogilvie},
  {Pringle}, \& {Tout}}]{bat00a}
{Bate}, M.~R., {Bonnell}, I.~A., {Clarke}, C.~J., {Lubow}, S.~H., {Ogilvie},
  G.~I., {Pringle}, J.~E., \& {Tout}, C.~A. 2000, \mnras, 317, 773

\bibitem[{{Bate} {et~al.}(1995){Bate}, {Bonnell}, \& {Price}}]{bat95}
{Bate}, M.~R., {Bonnell}, I.~A., \& {Price}, N.~M. 1995, \mnras, 277, 362

\bibitem[{{Binney} \& {Tremaine}(1987)}]{bin87}
{Binney}, J., \& {Tremaine}, S. 1987, {Galactic dynamics} (Princeton, NJ,
  Princeton University Press, 1987, 747 p.)

\bibitem[{{Boffin} {et~al.}(1998){Boffin}, {Watkins}, {Bhattal}, {Francis}, \&
  {Whitworth}}]{bof98}
{Boffin}, H.~M.~J., {Watkins}, S.~J., {Bhattal}, A.~S., {Francis}, N., \&
  {Whitworth}, A.~P. 1998, \mnras, 300, 1189

\bibitem[{{Bonnell} \& {Bate}(2002)}]{bon02}
{Bonnell}, I.~A., \& {Bate}, M.~R. 2002, \mnras, 336, 659

\bibitem[{{Bonnell} \& {Bate}(2005)}]{bon05}
---. 2005, \mnras, 362, 915

\bibitem[{{Bonnell} {et~al.}(2003){Bonnell}, {Bate}, \& {Vine}}]{bon03}
{Bonnell}, I.~A., {Bate}, M.~R., \& {Vine}, S.~G. 2003, \mnras, 343, 413

\bibitem[{{Bonnell} {et~al.}(1998){Bonnell}, {Bate}, \& {Zinnecker}}]{bon98a}
{Bonnell}, I.~A., {Bate}, M.~R., \& {Zinnecker}, H. 1998, \mnras, 298, 93

\bibitem[{{Clarke} {et~al.}(2000){Clarke}, {Bonnell}, \& {Hillenbrand}}]{cla00}
{Clarke}, C.~J., {Bonnell}, I.~A., \& {Hillenbrand}, L.~A. 2000, in Protostars
  and Planets IV, ed. V. Mannings, A. Boss, \& S. Russell (Tucson: Univ.
  Arizona Press), 151

\bibitem[{{Clarke} \& {Pringle}(1991)}]{cla91}
{Clarke}, C.~J., \& {Pringle}, J.~E. 1991, \mnras, 249, 584

\bibitem[{{Dolag} {et~al.}(2005){Dolag}, {Vazza}, {Brunetti}, \&
  {Tormen}}]{dol05}
{Dolag}, K., {Vazza}, F., {Brunetti}, G., \& {Tormen}, G. 2005, \mnras, 364,
  753

\bibitem[{{Gammie}(2001)}]{gam01}
{Gammie}, C.~F. 2001, \apj, 553, 174

\bibitem[{{Garc{\'{\i}}a} \& {Mermilliod}(2001)}]{gar01}
{Garc{\'{\i}}a}, B., \& {Mermilliod}, J.~C. 2001, \aap, 368, 122

\bibitem[{{Hall}(1997)}]{hal97}
{Hall}, S.~M. 1997, \mnras, 287, 148

\bibitem[{{Hall} {et~al.}(1996){Hall}, {Clarke}, \& {Pringle}}]{hal96}
{Hall}, S.~M., {Clarke}, C.~J., \& {Pringle}, J.~E. 1996, \mnras, 278, 303

\bibitem[{{Heller}(1995)}]{hel95}
{Heller}, C.~H. 1995, \apj, 455, 252

\bibitem[{{Hillenbrand} \& {Hartmann}(1998)}]{hil98}
{Hillenbrand}, L.~A., \& {Hartmann}, L.~W. 1998, \apj, 492, 540

\bibitem[{{Krumholz} {et~al.}(2005{\natexlab{a}}){Krumholz}, {Klein}, \&
  {McKee}}]{kru05d}
{Krumholz}, M.~R., {Klein}, R.~I., \& {McKee}, C.~F. 2005{\natexlab{a}}, in
  Protostars and Planets V, ed. B. Reipurth, D. Jewitt, \& K. Keil (Tucson:
  Univ. Arizona Press), 8271

\bibitem[{{Krumholz} {et~al.}(2005{\natexlab{b}}){Krumholz}, {McKee}, \&
  {Klein}}]{kru05b}
{Krumholz}, M.~R., {McKee}, C.~F., \& {Klein}, R.~I. 2005{\natexlab{b}}, \apjl,
  618, L33

\bibitem[{{Larwood}(1997)}]{lar97}
{Larwood}, J.~D. 1997, \mnras, 290, 490

\bibitem[{{McKee} \& {Tan}(2003)}]{mck03}
{McKee}, C.~F., \& {Tan}, J.~C. 2003, \apj, 585, 850

\bibitem[{{Monaghan}(1997)}]{mon97}
{Monaghan}, J.~J. 1997, Journal of Computational Physics, 136, 843

\bibitem[{{Monaghan}(2002)}]{mon02}
---. 2002, \mnras, 335, 843

\bibitem[{{Monaghan} \& {Gingold}(1983)}]{mon83}
{Monaghan}, J.~J., \& {Gingold}, R.~A. 1983, Journal of Computational Physics,
  52, 374

\bibitem[{{Morris} \& {Monaghan}(1997)}]{mor97}
{Morris}, J.~P., \& {Monaghan}, J.~J. 1997, Journal of Computational Physics,
  136, 41

\bibitem[{{Ostriker}(1994)}]{ost94}
{Ostriker}, E.~C. 1994, \apj, 424, 292

\bibitem[{{Pfalzner}(2004)}]{pfa04}
{Pfalzner}, S. 2004, \apj, 602, 356

\bibitem[{{Pfalzner} {et~al.}(2005){Pfalzner}, {Vogel}, {Scharw{\"a}chter}, \&
  {Olczak}}]{pfa05}
{Pfalzner}, S., {Vogel}, P., {Scharw{\"a}chter}, J., \& {Olczak}, C. 2005,
  \aap, 437, 967

\bibitem[{{Reipurth} {et~al.}(1999){Reipurth}, {Yu}, {Rodr{\'{\i}}guez},
  {Heathcote}, \& {Bally}}]{rei99}
{Reipurth}, B., {Yu}, K.~C., {Rodr{\'{\i}}guez}, L.~F., {Heathcote}, S., \&
  {Bally}, J. 1999, \aap, 352, L83

\bibitem[{{Rice} {et~al.}(2005){Rice}, {Lodato}, \& {Armitage}}]{ric05}
{Rice}, W.~K.~M., {Lodato}, G., \& {Armitage}, P.~J. 2005, \mnras, 364, L56

\bibitem[{{Ruffert}(1996)}]{ruf96}
{Ruffert}, M. 1996, \aap, 311, 817

\bibitem[{{Springel}(2005)}]{spr05}
{Springel}, V. 2005, \mnras, 364, 1105

\bibitem[{{Springel} \& {Hernquist}(2002)}]{spr02}
{Springel}, V., \& {Hernquist}, L. 2002, \mnras, 333, 649

\bibitem[{{Stahler} {et~al.}(2000){Stahler}, {Palla}, \& {Ho}}]{sta00}
{Stahler}, S.~W., {Palla}, F., \& {Ho}, P.~T.~P. 2000, in Protostars and
  Planets IV, ed. V. Mannings, A. Boss, \& S. Russell (Tucson: Univ. Arizona
  Press), 327

\bibitem[{{Watkins} {et~al.}(1998{\natexlab{a}}){Watkins}, {Bhattal}, {Boffin},
  {Francis}, \& {Whitworth}}]{wat98}
{Watkins}, S.~J., {Bhattal}, A.~S., {Boffin}, H.~M.~J., {Francis}, N., \&
  {Whitworth}, A.~P. 1998{\natexlab{a}}, \mnras, 300, 1205

\bibitem[{{Watkins} {et~al.}(1998{\natexlab{b}}){Watkins}, {Bhattal}, {Boffin},
  {Francis}, \& {Whitworth}}]{wat98a}
---. 1998{\natexlab{b}}, \mnras, 300, 1214

\bibitem[{{Wolfire} \& {Cassinelli}(1987)}]{wol87}
{Wolfire}, M.~G., \& {Cassinelli}, J.~P. 1987, \apj, 319, 850

\bibitem[{{Yorke} \& {Bodenheimer}(1999)}]{yor99}
{Yorke}, H.~W., \& {Bodenheimer}, P. 1999, \apj, 525, 330

\bibitem[{{Yorke} \& {Sonnhalter}(2002)}]{yor02}
{Yorke}, H.~W., \& {Sonnhalter}, C. 2002, \apj, 569, 846

\end{thebibliography}

\end{document}